\documentstyle[aps,preprint]{revtex}
\tighten
\begin{document}
\draft
\title{
  Finite-range-scaling analysis of metastability in an Ising model
  with long-range interactions
}
\author{Bryan M. Gorman and Per Arne Rikvold}
\address{
  Supercomputer Computations Research Institute,
  Department of Physics,
  and Center for Materials Research and Technology,
  Florida State University, Tallahassee, Florida 32306-4052
}
\author{M. A. Novotny}
\address{
  Supercomputer Computations Research Institute,
  Florida State University, Tallahassee, Florida 32306-4052
}

\date{\today}
\maketitle
\begin{abstract}
We apply both a scalar field theory and a recently developed
transfer-matrix method to study the stationary properties of
metastability in a two-state model with weak, long-range interactions:
the $N$$\times$$\infty$ quasi-one-dimensional Ising model.  Using the
field theory, we find the analytic continuation $\tilde f$ of the free
energy across the first-order transition, assuming that the system
escapes the metastable state by nucleation of noninteracting droplets.
We find that corrections to the field-dependence are substantial, and
by solving the Euler-Lagrange equation for the model numerically, we
have verified the form of the free-energy cost of nucleation,
including the first correction.  In the transfer-matrix method we
associate with subdominant eigenvectors of the transfer matrix a
complex-valued ``constrained'' free-energy density $f_\alpha$ computed
directly from the matrix.  For the eigenvector with an associated
magnetization most strongly opposed to the applied magnetic field,
$f_\alpha$ exhibits finite-range scaling behavior in agreement with
$\tilde f$ over a wide range of temperatures and fields, extending
nearly to the classical spinodal.  Some implications of these results
for numerical studies of metastability are discussed.
\end{abstract}

\section{Introduction}
\typeout{Introduction}
\label{sec-intro}

Determining the stationary properties of metastable states from the
standpoint of statistical mechanics has been the topic of many studies
over the last three decades.  (For reviews, see {\it e.g.}\ Refs.\
\cite{Gunt83i,Bind87}.)  Some early treatments of this problem have
shown that traditional methods of equilibrium statistical mechanics in
a more generalized form might be applicable to metastable states.  In
a study of the analytic properties of the free energy at the
condensation point, Langer \cite{Lang67} conjectured that the
imaginary part of the free energy $\widetilde F$ analytically
continued from the equilibrium phase across the first-order phase
transition may be associated with the decay rate of the metastable
phase.  A dynamical investigation of thermally activated nucleation
\cite{Lang68,Lang69} showed that for a wide class of models, the decay
rate $\Gamma$ may be written as
\begin{equation}
\label{int-eq0}
\Gamma = \frac{\beta\kappa}\pi|{\rm Im}\widetilde F|\;,
\end{equation}
where $\beta$ is the inverse temperature $1/k_{\rm B}T$, and $\kappa$
is a kinetic prefactor that contains all dependence on the dynamics.

Subsequently there have been a number of theoretical and numerical
studies of metastable decay.  Binder and collaborators
\cite{Bind73a,Bind73b,Bind74} developed a scaling theory based on a
proposed definition of metastable states in terms of a nonequilibrium
relaxation function and tested this theory by Monte Carlo simulation
on the 2D Ising model.  Schulman and coworkers studied by various
methods metastability in the 2D Ising model \cite{McCr78}, in the 1D
Kac model with algebraically decaying interactions \cite{McCr80}, in
the Curie-Weiss model \cite{Newm80}, and in a dropletlike ``urn''
model \cite{Roep84}.  In addition, B\"uttiker and Landauer
\cite{Buet79,Buet81} studied nucleation in the overdamped 1D
sine-Gordon chain, and Klein and Unger \cite{Klei83,Unge84} used a
$\phi^3$ field theory to study systems with long-range interactions in
arbitrary dimensions near the classical spinodal.  Other studies of
nucleation in short-range systems also exist in the literature,
including field-theoretical studies \cite{Call77,Guen80,Cott93},
series expansions \cite{Baxt79,Bake80,Lowe80,Enti80}, and exact
diagonalization studies \cite{Hame83}.  Each of the above studies
supported the validity of Langer's treatment.  Recently Gaveau and
Schulman \cite{Gave89} have determined a rigorous upper bound for
$\Gamma$ for a larger class of models than that considered by Langer,
and have used it not only to explain why Eq.\ (\ref{int-eq0}) is
usually valid, but also to provide an example in which it can give
misleading results.

A study of the analytic properties of transfer-matrix eigenvalues led
to a conjecture \cite{Newm77} that there is a correspondence between
the analytically continued free-energy density $\tilde f$ and the
analytic continuation of the dominant eigenvalue of the transfer
matrix.  This conjecture has been supported by subsequent numerical
studies \cite{McCr78,Priv82a,Priv82b} of the 2D Ising model.  In a
similar study \cite{Novo86} of an Ising model with $N$$\times$$\infty$
cylindrical geometry in which the interaction range is linear in the
cross section, evidence from the transfer-matrix eigenvalue spectrum
was found for the emergence of a classical spinodal in the limit of
weak, long-range interactions.  In the present work we apply a
recently developed ``constrained-transfer-matrix'' (CTM) method
\cite{Rikv89,Rikv91a,Rikv92a,Guen93} to study the properties of the
metastable phase of this quasi-one-dimensional Ising (Q1DI) model.
Since the transfer matrix for the $N$$\times$$\infty$ Q1DI model has a
rank of $N$+1 \cite{Novo86}, instead of the typical value of $2^N$ for
short-range Ising systems, it is relatively easy to study large
systems by transfer-matrix techniques.  Using the CTM method, we
obtain a ``constrained'' free-energy density associated with the
metastable phase.  Our main focus here will be to relate the
finite-size (or equivalently for this model, finite-{\em range})
scaling behavior of the constrained free energy to the scaling
behavior of the analytic continuation of the free energy for an
equivalent one-dimensional long-range field-theoretical model.  This
work represents an extension to metastable states of our recent study
\cite{Rikv93} of critical equilibrium finite-range scaling in the Q1DI
model.

In studies of condensation it is often useful to consider the behavior
of long-range models.  One reason for this is that long-range models
share many of the equilibrium and metastable properties of mean-field
models, which are often exactly soluble.  In addition, as with
equilibrium states near the critical point, the metastable states of
these models near the classical spinodal may be studied using a simple
field-theoretical Hamiltonian \cite{Klei83,Unge84}.  Further, the
behavior of many physical systems, such as coherent metal-hydrogen
systems \cite{Wagn74}, superfluid $^3$He \cite{Legg75}, long-chain
polymer mixtures \cite{DeGe79,Bind84}, materials exhibiting elastic
phase transitions \cite{Cowl80}, superconductors \cite{Schr83}, and
ferroelectrics with long-range strain fields \cite{Litt86,Chan89},
indeed are well described by long-range models with suitably chosen
order parameters.  Many treatments of classical metastability for
these models are reported in the literature, using many different
approaches: the Fokker-Planck equation \cite{Newm80,Grif66,Paul89},
the properties of the nonequilibrium relaxation function
\cite{Bind73a,Paul89}, renormalization group analysis \cite{Gunt78},
and Monte Carlo simulation \cite{Grif66,Paul89,Ray90,Ray91}, in
addition to analytic continuation of the free energy
\cite{Newm80,Klei83,Unge84}.  The approach of the present work is to
calculate numerically from the transfer matrix a constrained
free-energy density \cite{Rikv89,Rikv91a} associated with the
classical metastable state, and to determine its scaling behavior in
weak fields away from the critical point, and near the classical
spinodal.  We demonstrate that this quantity is closely related to the
analytically continued free-energy density.

The remainder of this paper is outlined as follows.  In Sec.\
\ref{sec-model} we derive the Euler-Lagrange equation for the order
parameter in the Q1DI model, and we identify the solutions
corresponding to the equilibrium and metastable phases, as well as the
most likely fluctuation through which a Q1DI system in the metastable
phase can decay into the equilibrium phase.  In Sec.\
\ref{sec-analytic} we obtain a mapping of the Q1DI model near the
classical spinodal to a one-dimensional $\phi^3$ field theory.  Using
both the solution to the Euler-Lagrange equation and the field theory,
we perform the analytic continuation of the free energy for the model
across the first-order phase transition, obtaining an exact expression
for its imaginary part near the classical spinodal. In Sec.\
\ref{sec-method} we give a brief overview of the CTM method, which is
based on the notion of ``constrained'' joint probability densities
\cite{Rikv89,Rikv91a}, and we show how the symmetries of the Q1DI
model \cite{Novo86} are used to simplify the calculation.  In Sec.\
\ref{sec-results} we verify by numerical integration of the
Euler-Lagrange equation the temperature and field dependence of the
the free-energy cost of nucleation, including the first field
correction. We also present our numerical transfer-matrix results and
compare them with finite-range-scaling predictions for $\tilde f$
based on the field-theoretical results of Sec. \ref{sec-analytic}.  In
Sec.\ \ref{sec-disc} we summarize our conclusions, and we discuss the
implications of the results to the applicability of the CTM method to
studies of metastability.

\section{The Q1DI Model}
\typeout{The Q1DI Model}
\label{sec-model}

The quasi-one-dimensional Ising, or Q1DI, model \cite{Novo86} is a
one-dimensional chain of $L$ subsystems, or layers, each of which
contains $N$ Ising spins $s_{i,n}$=$\pm1$, where the index
$i$=$1,\ldots,L$ runs over the layers, and the index $n$=$1,\ldots,N$
runs over the spins in a given layer.  Each spin interacts
ferromagnetically with each of the $2N$ spins in the adjacent layers
with a coupling $J/N$$>$0 and with the other $N$$-1$ spins of its own
layer with a coupling $J'/N$.  Each spin also interacts with an
external magnetic field $H$.  The Hamiltonian is explicitly
\begin{equation}
{\cal H} = -\sum_{i=1}^L\left(
\frac JN\sum_{n=1}^N\sum_{n'=1}^Ns_{i,n}s_{i+1,n'}
+\frac{J'}N\sum_{n=1}^{N-1}\sum_{n'=n+1}^N
s_{i,n}s_{i,n'}+H\sum_{n=1}^Ns_{i,n}\right)\;,\label{mod-eq0}
\end{equation}
where periodic boundary conditions are imposed, that is ($\forall i,n$)
$s_{L+i,n}$=$s_{i,n}$.  The sums over $n$ and $n'$ in Eq.\
(\ref{mod-eq0}) may be performed directly, so this Hamiltonian may be
expressed in terms of a discrete field $\phi$ of layer magnetization
densities $\phi_i$=$N^{-1}\sum_{n=1}^Ns_{i,n}$ as
\begin{equation}
{\cal H}[\phi] = -N\sum_{i=1}^L\left[J\phi_i\phi_{i+1}
+\case12J'(\phi_i^2-N^{-1})+H\phi_i\right]\;.\label{mod-eq1}
\end{equation}
The partition function may then be written as
\begin{equation}
Z = \sum_\phi g[\phi]e^{-\beta{\cal H}[\phi]}\;,\label{mod-eq2a}
\end{equation}
where
\begin{equation}
g[\phi] = \prod_{i=1}^L{N\choose{\case12N(1+\phi_i)}}\label{mod-eq2b}
\end{equation}
denotes the multiplicity of spin configurations giving rise to $\phi$.
In the limit of large $N$, the probability density of $\phi$ is
sharply peaked around its maximum, so if we define a
free-energy-density functional as
\begin{equation}
{\cal F}[\phi] = (NL)^{-1}\left(
{\cal H}[\phi]-\beta^{-1}\ln(g[\phi])\right)\;,\label{mod-eq3}
\end{equation}
then the free-energy density is
\begin{equation}
f_{\rm Q1DI} = -(\beta NL)^{-1}\ln Z \approx \min_\phi{\cal F}[\phi]\;.
\label{mod-eq4}
\end{equation}
Applying Stirling's approximation to Eq.\ (\ref{mod-eq2b}), combining
the result with Eqs.\ (\ref{mod-eq1}) and (\ref{mod-eq3}), and
removing constant terms, we have for the Q1DI model
\begin{equation}
{\cal F}[\phi] = \frac1L\sum_{i=1}^L
\left[\case12J(\phi_{i+1}-\phi_i)^2+f_{\rm CW}(\phi_i)\right]\;,
\label{mod-eq5}
\end{equation}
where
\begin{equation}
f_{\rm CW}(\phi_i) = -\case12\beta_{\rm c}^{-1}\phi_i^2-H\phi_i
+\beta^{-1}\left(\frac{1+\phi_i}2\ln\frac{1+\phi_i}2
+\frac{1-\phi_i}2\ln\frac{1-\phi_i}2\right)
\label{mod-eq5a}
\end{equation}
is the free-energy density of a Curie-Weiss ferromagnet with a
critical temperature
\begin{equation}
\beta_{\rm c}^{-1}=2J+J'.
\label{mod-eq5b}
\end{equation}
Corrections to $f_{\rm CW}$ due to finite $N$ will be briefly considered
in Sec.\ \ref{sec-disc}.

In Ref.\ \cite{Rikv93} we showed, both analytically and numerically,
that the critical equilibrium $N$-scaling relations for the Q1DI model
are those of an $N$$\times$$\infty$ cylindrical Ising model above its
upper critical dimension, which can be obtained from a standard
one-dimensional $\phi^4$ field theory by a single rescaling of the
correlation length.  Here we extend our study of the scaling behavior
to metastable states, where $\beta$$>$$\beta_{\rm c}$ and $H$$\ne$0,
by studying the scaling behavior of the analytic continuation of the
free energy across the first-order phase transition.  To do this, we
must first calculate the free-energy density for the equilibrium and
metastable phases of the model, as well as the height of the
free-energy barrier between them.  For large $N$, we can treat each
$\phi_i$ as a continuous variable.  In this case, extremizing ${\cal
F}[\phi]$ with respect to $\phi$ in the $L$-cube $[-1,1]^L$, we obtain
a system of $L$ coupled nonlinear equations:
\begin{equation}
L\frac{\partial{\cal F}}{\partial\phi_i} =
-J(\phi_{i+1}-2\phi_i+\phi_{i-1})
+\frac{\partial f_{\rm CW}(\phi_i)}{\partial\phi_i} = 0\;.
\label{mod-eq6}
\end{equation}

We can gain some understanding of the solutions to Eq.\
(\ref{mod-eq6}) by expressing the equation in differential form.  We
define a function $\phi(r)$ to be continuous in a dimensionless
longitudinal coordinate $r$, and we force $\phi(r)$=$\phi_i$ at
integer values $r$=$i$.  By Taylor's theorem, if $i\le r\le i+1$, then
$(\phi_{i+1}-\phi_i)^2=(\nabla\phi(r))^2+O(\nabla\phi\nabla^2\phi)$.
Therefore, if $\phi(r)$ does not vary too rapidly over the length of
the chain, then the free-energy density is well approximated by the
Ginzburg-Landau-Wilson form:
\begin{equation}
{\cal F}[\phi] = \frac1L\int_0^Ldr
\left[\case12J(\nabla\phi)^2+f_{\rm CW}(\phi(r))\right]\;,
\label{mod-eq7}
\end{equation}
which represents an effective interaction range
\begin{equation}
R_N=\sqrt{\beta_{\rm c}J}N\;,
\end{equation}
as measured in sites.  Note, however, that $R_N$ is finite as measured
in layers.  In the critical region, this mapping to the continuum
field theory agrees with the formulation in Ref.\ \cite{Rikv93}.  The
resulting Euler-Lagrange equation for stationary solutions is the
continuous analogue to Eq.\ (\ref{mod-eq6}):
\begin{equation}
L\frac{\delta{\cal F}[\phi]}{\delta\phi(r)}
= -J\frac{d^2\phi}{dr^2}
+\left.\frac{\partial f_{\rm CW}}{\partial\phi}\right|_{\phi(r)}
= 0\;,\label{mod-eq8a}
\end{equation}
which can be integrated once to give
\begin{equation}
\case12J\left(\frac{d\phi}{dr}\right)^2 = f_{\rm CW}(\phi(r))-C\;,
\label{mod-eq8b}
\end{equation}
where $C$ is a constant.

Equation (\ref{mod-eq8a}) may be interpreted as the equation of motion
for a particle of mass $J$ moving in a potential $-f_{\rm
CW}(\phi(r))$, where $\phi(r)$ represents the displacement of the
particle at time $r$.  Detailed descriptions of the solution from this
point of view can be found in Refs.\ \cite{Lang67,Unge84}.  In
determining the solutions, we assume that $\phi(r)$ is continuous.
Since $f_{\rm CW}$ is continuous on the interval $[-1,1]$ and
continuously differentiable on $(-1,1)$, $\phi(r)$ is twice
continuously differentiable on $(-1,1)$.  The boundedness of $\phi$
requires that if the LHS of Eq.\ (\ref{mod-eq8b}) is nonzero for all
$r$, it must approach zero as $|r|$$\rightarrow$$\infty$, since $\phi$
would in that case be monotone, while the nonnegativity of the LHS of
Eq.\ (\ref{mod-eq8b}) requires that ($\forall r$) $f_{\rm
CW}(\phi(r))$$\geq$$C$.  The first condition on the solution is
therefore that the range of $\phi$ is an interval $I$$\subset$$[-1,1]$
for which $\inf(f_{\rm CW}(I))$=$C$.  Further, we must have as a
second condition that $f_{\rm CW}(\phi)$=$C$ at the endpoints of $I$.
To see this, assume on the contrary that $f_{\rm CW}(\phi')$$>$$C$ for
one of the endpoints $\phi'$.  For some $r$ we would then have
$\phi(r)$=$\phi'$ and $|d\phi/dr|$$>$0 by Eq.\ (\ref{mod-eq8b}),
forcing $\phi(r)$ to violate the first condition by taking values
outside the interval $I$.

By varying the constant $C$ we find three types of solutions
satisfying the above constraints.  These are illustrated in Fig.\
\ref{fig1}.  The first type (I) is a completely uniform solution, in
which $d\phi/dr$=$d^2\phi/dr^2$=0 for all $r$.  By Eq.\
(\ref{mod-eq8a}), these solutions exist only if the constant value of
$\phi(r)$ is an extremum of $f_{\rm CW}$. There can be as many as
three solutions of this type, and they are marked on the free-energy
sketches in Fig.\ \ref{fig1} as diamonds.  The second type of solution
(II) is a nonuniform ``droplet'' or ``interface'' solution, in which
one of the endpoints of $I$ is the locus of a local minimum $\phi_{\rm
ms}$ of $f_{\rm CW}$ that is not a unique global minimum.  If such a
minimum exists, there is only one solution of this type.  The vertical
lines descending from the free-energy sketches in Fig.\ \ref{fig1}
mark the endpoints of the range of $\phi(r)$ for this solution, and
sketches of the solution in real space are drawn below.  If the
type-II solution exists, then a band of solutions of a third type
(III) also exists.  These solutions are oscillatory in space and are
marked on the free-energy sketches in Fig.\ \ref{fig1} as hashed
regions.  They are not considered in the following discussion since
the free-energy densities associated with these solutions are the
highest of those for the three types.

By considering the solutions of types I and II, we easily identify the
equilibrium and metastable states as those with the two lowest values
of $\cal F$.  Both are uniform (type-I) solutions,
$\phi(r)$=$\phi_{\rm eq}$ and $\phi(r)$=$\phi_{\rm ms}$, respectively,
and thus by Eq.\ (\ref{mod-eq6}) have ${\cal F}[\phi]$=$f_{\rm
CW}(\phi_{\rm eq})$ and ${\cal F}[\phi]$=$f_{\rm CW}(\phi_{\rm ms})$,
respectively.  The metastable state exists at temperatures and fields
for which $f_{\rm CW}$ exhibits two minima.  For $\beta$$>$$\beta_{\rm
c}$, this is true if $\phi_{\rm ms}$ does not coincide with a point of
inflection of $f_{\rm CW}$.  The locus of the inflection is given by
$\phi=\pm\sqrt{1-\beta_{\rm c}/\beta}$, so the metastable state
exists if $\phi_{\rm ms}^2>1-\beta_{\rm c}/\beta$.  Combining this
constraint with Eq.\ (\ref{mod-eq6}) and with the uniformity of
$\phi$, we obtain the classical (Curie-Weiss) spinodal field $H_{\rm
s}$ \cite{Newm80,Paul89}, which defines the limit of metastability:
\begin{equation}
H_{\rm s} = -\beta_{\rm c}^{-1}\phi_{\rm s}
+\beta^{-1}\tanh^{-1}\phi_{\rm s}\;,
\label{mod-eq9a}
\end{equation}
where
\begin{equation}
\phi_{\rm s} = -({\rm sgn}H)\sqrt{1-\beta_{\rm c}/\beta}\;.
\label{mod-eq9b}
\end{equation}
The solution with the next-lowest value of $\cal F$ is the type-II
solution.  This solution represents the fluctuation of lowest free
energy through which the metastable phase can decay into the
equilibrium phase (for $H$$\ne$0), or through which a system can pass
between two coexisting phases (for $H$=0).

For 0$<$$|H|$$<$$|H_{\rm s}|$, we can estimate the free-energy cost
$\Delta F$ of this fluctuation by numerically solving the coupled
nonlinear equations defined by Eq.\ (\ref{mod-eq6}) for an $L$-layer
system, allowing the magnetization densities to vary continuously.  In
Sec.\ \ref{sec-results} the numerical integration procedure is
outlined and its results for $\Delta F$ are compared with the analytic
result from a field-theoretical Hamiltonian near the spinodal and with
our transfer-matrix results.

\section{Analytic Continuation of $F$}
\typeout{Analytic Continuation of $F$}
\label{sec-analytic}

For the Q1DI model with $\beta$$>$$\beta_{\rm c}$, the free energy $F$
in the limit $N$$\rightarrow$$\infty$ is not everywhere analytic, but
rather exhibits a discontinuous first derivative with respect to $H$
at $H$=0.  However, if we vary $H$ continuously through $H$=0, an
analytic continuation $\tilde f$ of the free-energy density across
$H$=0 does exist as an analogue of Eq.\ (\ref{mod-eq4}), where the
partition function is constrained to configurations that do not allow
the system to reach equilibrium.  This continuation is the minimum of
${\cal F}$ that coincides with $f_{\rm Q1DI}$ at $H$=0, but increases
as $H$ departs from zero.  When $H$ reaches the spinodal field, the
metastable minimum vanishes.  At this point the continuation has a
branch point and becomes complex as $|H|$ is increased further.

We determine the leading behavior of the analytic continuation of $F$
in the region of the spinodal ($\phi$=$\phi_{\rm s}$, $H$=$H_{\rm s}$)
as follows.  Taking the continuum limit used in the previous section,
we write $\phi(r) = \phi_{\rm s}+v(r)$ and $H = H_{\rm s}+\lambda$,
and we expand Eq.\ (\ref{mod-eq7}) for the free-energy-density
functional near the spinodal to third order:
\begin{equation}
{\cal F}[T,H,\phi] =
{\cal F}[T,H_{\rm s},\phi_{\rm s}]+\Delta{\cal F}\;,
\end{equation}
with
\begin{eqnarray}
\Delta{\cal F} &=& \frac1L\int_0^Ldr\left[
\case12J(\nabla v)^2+\lambda\left.
\frac{\partial f_{\rm CW}}{\partial H}\right|_{\phi_{\rm s}}
+\lambda v\left.
\frac{\partial^2f_{\rm CW}}{\partial\phi\partial H}
\right|_{\phi_{\rm s}}
+\case16v^3\left.
\frac{\partial^3f_{\rm CW}}{\partial\phi^3}\right|_{\phi_{\rm s}}
+O(v^4)\right]\nonumber\\
&=& -\lambda\phi_{\rm s}
+\frac1L\int_0^Ldr\left[\case12J(\nabla v)^2-\lambda v
+\case13\alpha v^3+O(v^4)\right]\;,\label{acf-eq0}
\end{eqnarray}
where we find
\begin{equation}
\alpha = \frac{\beta^{-1}\phi_{\rm s}}{(1-\phi_{\rm s}^2)^2}
= -({\rm sgn}H)\frac\beta{\beta_{\rm c}^2}
\sqrt{1-\frac{\beta_{\rm c}}\beta}
\label{acf-eq0a}
\end{equation}
using the explicit form of Eq.\ (\ref{mod-eq5a}) for $f_{\rm CW}$ and
Eq.\ (\ref{mod-eq9b}).  In Eq.\ (\ref{acf-eq0}) we have shown only the
derivatives that are not identically zero. (Since $f_{\rm CW}$ is
stationary and has an inflection at $\phi$=$\phi_{\rm s}$, the first
and second $\phi$-derivatives are among those not shown.)  From the
above expansion, the requirement that $\cal F$ is stationary gives the
Euler-Lagrange equation:
\begin{equation}
-J\nabla^2v-\lambda+\alpha v^2+O(v^3) = 0\;.\label{acf-eq1}
\end{equation}

Since the critical fluctuation consists of a single droplet, the
free-energy cost of the fluctuation is not extensive in $L$.  We take
the position of the droplet core to be $r$=0 in the following
discussion.  Since this fluctuation is local, the solution to Eq.\
(\ref{acf-eq1}) must be asymptotically uniform in the limit
$L$$\rightarrow$$\infty$, that is $v(r)$$\rightarrow$constant as
$|r|$$\rightarrow$$\infty$.  From Eq.\ (\ref{acf-eq1}) we thus have
asymptotically
\begin{equation}
v \rightarrow v_0 = (\lambda/\alpha)^{1/2}+O(\lambda)\label{acf-eq2}
\end{equation}
with the sign restriction ${\rm sgn}v_0$=${\rm sgn}\alpha$ if
$|H|$$<$$|H_{\rm s}|$.  Changing variables to $u(r) = v(r)-v_0$, the
Euler-Lagrange equation becomes
\begin{equation}
-J\nabla^2u+2\alpha v_0u+\alpha u^2+O(u^3) = 0\;.\label{acf-eq3}
\end{equation}
In the limit $|r|$$\rightarrow$$\infty$ we may neglect terms of
O($u^2$).  We thus obtain
\begin{equation}
u(r) \sim \exp(-r/\xi_{\rm r})\;,\label{acf-eq4}
\end{equation}
where
\begin{equation}
\xi_{\rm r} = \sqrt{\case J2}|\alpha\lambda|^{-1/4}+O(\lambda^{1/4})
\end{equation}
is the relaxation length of the fluctuation.  The power-law dependence
on $\lambda$ of the order parameter $v_0$ and the length scale
$\xi_{\rm r}$ is indicative of a critical point at the spinodal, as
was pointed out previously in Refs.\ \cite{Klei83,Unge84,Gunt78}.  We
can use the change of variables above to separate the $L$-extensive
part of the integral in Eq.\ (\ref{acf-eq0}):
\begin{eqnarray}
\Delta{\cal F} &=& -\lambda\phi_{\rm s}
+\frac1L\int_0^Ldr\left[\case12J(\nabla u)^2
+\alpha(-\case23v_0^3+v_0u^2+\case13 u^3)+O(v^4)\right]\nonumber\\
&=& -\lambda\phi_{\rm s}-\case23\alpha v_0^3+O(v_0^4)
-\frac1L\int_0^Ldr\left[\case16\alpha u^3+O(u^4)\right]\;,
\label{acf-eq5}
\end{eqnarray}
where we have integrated $(\nabla u)^2$ by parts and used Eqs.\
(\ref{acf-eq2}) and (\ref{acf-eq3}).

If $|H|$$>$$|H_{\rm s}|$, then $v_0$ is purely imaginary.  For the
uniform solutions of the Euler-Lagrange equation, $v$=$\pm i|v_0|$,
$\Delta{\cal F}$ has an imaginary part given, to this order of
approximation, by the second term in Eq.\ (\ref{acf-eq5}).  Combining
this with Eq.\ (\ref{acf-eq2}) and noting that $v_0^4$ is real, we
have
\begin{equation}
{\rm Im}\tilde f = {\rm Im}(\Delta{\cal F}) =
\pm\case23|\alpha|^{-1/2}|\lambda|^{3/2}+O(\lambda^{5/2})\;.
\label{acf-eq6}
\end{equation}
This expression is exactly the result for a Curie-Weiss mean-field
model \cite{Newm80,Paul89}.

On the other hand, if $|H|$$<$$|H_{\rm s}|$, then $\Delta{\cal F}$ is
purely real, and the first three terms of Eq.\ (\ref{acf-eq5}) give
the contribution of the uniform metastable background, whereas the
integral term gives the contribution of the critical droplet.  Taking
the limit $L$$\rightarrow$$\infty$, we insert the explicit solution to
Eq.\ (\ref{acf-eq1}) \cite{Buet81,Klei83,Unge84},
\begin{equation}
v_1(r) = v_0\left[1-3{\rm sech}^2(r/2\xi_{\rm r})\right]\;,
\label{acf-eq7}
\end{equation}
into the integrand in Eq.\ (\ref{acf-eq5}) to find the free-energy
cost $\Delta F$ of forming the droplet:
\begin{eqnarray}
\Delta F &=&
\case{48}5N\xi_{\rm r}|\alpha|^{-1/2}|\lambda|^{3/2}
\left[1+O(\lambda^{1/2})\right]\nonumber\\
&=& \case{24}5N\sqrt{2J}|\alpha|^{-3/4}|\lambda|^{5/4}
+O(\lambda^{7/4})\;.\label{acf-eq8}
\end{eqnarray}
The first expression for $\Delta F$ in Eq.\ (\ref{acf-eq8})
illustrates the difference in the behavior of the analytically
continued free energy between long-range models, for which the length
scale for the critical droplet is a characteristic length $\xi_{\rm
r}$$\ll$$L$ \cite{Klei83,Unge84}, and mean-field models, for which the
only length scale for fluctuations is the length $L$ of the entire
system.  The volume-extensivity and field-dependence of the mean-field
result for $\Delta F$ \cite{Newm80,Paul89} can be recovered by
replacing $\xi_{\rm r}$ with $L$ in Eq.\ (\ref{acf-eq8}).

Note that if $|H|$$<$$|H_{\rm s}|$, the analytically continued free
energy $\widetilde F$ is real-valued only in the limit of infinite
interaction range.  However, we are interested in the scaling of the
free energy for systems with a long, but finite interaction range.
For such systems $\widetilde F$ is complex, but its imaginary part
approaches zero rapidly as $R_N$$\rightarrow$$\infty$.  We can
determine the scaling properties of the free energy by expanding the
partition function for the system under the constraint that all
fluctuations remain subcritical.  Following Refs.\
\cite{Lang67,Schu81}, we write
\begin{equation}
Z=Z_0+Z_1+Z'\;,\label{acf-eq9}
\end{equation}
where $Z_0$ is the contribution from the region of configuration space
around the local minimum of ${\cal F}$ corresponding to the metastable
background, $Z_1$ is the contribution from the region around the
saddle point corresponding to the critical droplet, and $Z'$
represents higher-order terms. The free energy must be extensive in
$L$, so we must also consider contributions of multiple critical
droplets.  If we assume that $Z'$ has contributions only from multiple
identical, noninteracting droplets, then by expanding $Z$ as a series
in $Z_1/Z_0$ we have \cite{Schu81}
\begin{equation}
Z = Z_0\exp(Z_1/Z_0)\;.\label{acf-eq10}
\end{equation}

For the stationary points of $\cal F$ identified above as the uniform
metastable background $v_0$ and the critical droplet $v_1$, we write
$v(r) = v_n(r)+\nu(r)$ and expand ${\cal F}$ to second order:
\begin{equation}
{\cal F}[v] \approx {\cal F}[v_n]+\delta^2{\cal F}\;,
\label{acf-eq11}
\end{equation}
where
\begin{eqnarray}
\delta^2{\cal F} &=&
\frac1L\int_0^Ldr\left[\case12J(\nabla\nu)^2+\case12\nu^2\left.
\frac{\partial^2f_{\rm CW}}{\partial\phi^2}\right|_{\phi_{\rm s}+v_n}
\right]\nonumber\\
&=& \frac1L\int_0^Ldr\left[\case12J(\nabla\nu)^2
+\left(\alpha v_n+O(v_n^2)\right)\nu^2\right]\;.\label{acf-eq12}
\end{eqnarray}
In Eq.\ (\ref{acf-eq12}) we have expanded the second term of the
integrand to first order in $v_n$ since the values of $v_n(r)$ are
small, and we have again used the fact that $f_{\rm CW}$ has an
inflection at $\phi$=$\phi_{\rm s}$.  We diagonalize the quadratic
form by a principal-axis transformation $\nu(r)$=$\sum_ja_j\nu_j$ to
the orthonormal set of eigenmodes of the Schr\"odinger equation
\begin{equation}
-\case12J\nabla^2\nu_j+\left(\alpha v_n+O(v_n^2)\right)\nu_j
= \omega_j\nu_j\;.
\label{acf-eq13}
\end{equation}
{}From Eqs.\ (\ref{acf-eq11})--(\ref{acf-eq13}) the resulting
contribution $Z_n$ to the partition function can thus be written as a
product of decoupled Gaussian integrals:
\begin{eqnarray}
Z_n &=& \int dv\exp(-\beta NL{\cal F}[v])\nonumber\\
&=& \exp(-\beta NL{\cal F}[v_n])
\prod_j\int da_j\exp(-\beta Na_j^2\omega_j)\;.
\label{acf-eq14}
\end{eqnarray}
Note that although the results that appear later in this section
are restricted to the region of the spinodal, Eq.\ (\ref{acf-eq14})
holds everywhere in the region of metastability.  It can easily be
seen from this and from Eq.\ (\ref{acf-eq10}) that in this region a
Boltzmann weight equal to $e^{-\beta\Delta F}$ appears in the
analytically continued free energy.

In order to determine the behavior of the analytically continued free
energy in the region of the spinodal, where $\Delta F$ is small, we
must further determine the eigenvalues of Eq.\ (\ref{acf-eq13}).  For
$n$=0 Eq.\ (\ref{acf-eq13}) is a free-particle equation with a
potential floor at $\alpha v_0$, whereas for $n$=1 it describes a
particle in a potential well described by $\alpha v_1(r)$.  There are
three localized eigenmodes for $n$=1 \cite{Buet81}.  They are
\begin{eqnarray}
\nu_0 &\propto&{\rm sech}^3(r/2\xi_{\rm r})\nonumber\\
\nu_1 &\propto&
{\rm sech}^2(r/2\xi_{\rm r}){\rm tanh}(r/2\xi_{\rm r})\nonumber\\
\nu_2 &\propto&
4{\rm sech}(r/2\xi_{\rm r})-5{\rm sech}^3(r/2\xi_{\rm r})\;,
\label{acf-eq15}
\end{eqnarray}
for which
\begin{eqnarray}
\omega_0 &=& -\case58J/\xi_{\rm r}^2
= -\case54|\alpha\lambda|^{1/2}+O(\lambda)\nonumber\\
\omega_1 &=& 0+O(\lambda)\nonumber\\
\omega_2 &=& \case38J/\xi_{\rm r}^2
= \case34|\alpha\lambda|^{1/2}+O(\lambda)\label{acf-eq16}
\end{eqnarray}
are the respective eigenvalues \cite{frsam-note1}.  The first
eigenmode $\nu_0$ is the only unstable mode of fluctuation and
corresponds to an increasing or decreasing magnetization at the
droplet core.  The second eigenmode $\nu_1$ is a translational mode,
and the third eigenmode $\nu_2$ corresponds to a widening or narrowing
of the droplet.  All other eigenmodes are unbounded and form a
continuous spectrum with $\omega_j$$\ge$$\case12J/\xi_{\rm r}^2$.

We shall consider the contributions of each localized eigenmode
individually.  For each $\omega_j$$>$0, the Gaussian integral of Eq.\
(\ref{acf-eq14}) is a well-defined function, $G(\omega_j) = (\pi/\beta
N\omega_j)^{1/2}$, but if $\omega_j$$\le$0, as is the case for $j$=0
and $j$=1, the Gaussian integral is divergent.  We handle each
divergence in a different way.  For $j$=0, we analytically continue
$G(\omega)$ from the half-plane defined by ${\rm Re}\omega$$>$0 to the
entire complex plane.  The result is purely imaginary and has a sign
ambiguity, since the continuation of $G$ has a branch cut on the
negative real axis.  For $j$=1, we use the fact that $\nu_1$ is the
normalized derivative of $v_1$ (Eq.\ (\ref{acf-eq7})) with respect to
$r$.  We can thus make a transformation of the integral to one over
$dr$, the Jacobian of which is simply $\|\nabla v_1\|$.  The resulting
contributions to $Z_1$ from the localized modes are
\begin{eqnarray}
\pm\frac i2\sqrt{\frac\pi{\beta N|\omega_0|}} &\;\;\;\;& (j=0)
\nonumber\\
\sqrt{\frac{\Delta F}{NJ}}L &\;\;\;\;& (j=1)\nonumber\\
\sqrt{\frac\pi{\beta N\omega_2}} &\;\;\;\;& (j=2)\;.
\label{acf-eq17}
\end{eqnarray}
The factor of $\frac12$ has been introduced in the $j$=0 result
\cite{frsam-note2} to account for the fact that the integral
inappropriately counts a divergence as one moves toward the metastable
minimum \cite{Schu81}.  The form of the $j$=1 result
\cite{frsam-note2} is quite general and follows from a simple virial
argument \cite{Call77,Guen80}.  We are then left with the
contributions to $Z_1/Z_0$ from the continuous spectra of eigenmodes.
Using a previous calculation of the ratio of the relevant eigenvalue
products in App.\ B of Ref.\ \cite{Buet81}, we obtain
\begin{equation}
\frac{\prod_{j=3}^\infty G(\omega_j)}
{\prod_{j=0}^\infty G(\omega_j^0)}
= \frac{\sqrt{30J|\omega_0|\omega_2}}{\xi_{\rm r}}
\left(\frac{\beta N}\pi\right)^{3/2}\;,
\label{acf-eq18}
\end{equation}
where $\omega_j^0$ denotes an eigenvalue of the free-particle
Schr\"odinger equation.  Combining this with Eqs.\ (\ref{acf-eq10}),
(\ref{acf-eq14}), (\ref{acf-eq16}), and (\ref{acf-eq17}), we have
\cite{frsam-note2}
\begin{eqnarray}
{\rm Im}\tilde f &=& \pm\sqrt{\frac{15}{2\pi\beta}}
\frac{(\Delta F)^{1/2}}{N\xi_{\rm r}}e^{-\beta\Delta F}\nonumber\\
&=& \pm12(\sqrt{2J}\pi\beta N)^{-1/2}
|\alpha|^{-1/8}|\lambda|^{7/8}\left[1+O(\lambda^{1/2})\right]
e^{-\beta\Delta F}\;.\label{acf-eq19}
\end{eqnarray}

Assuming that the dynamics of the system is governed by a
Fokker-Planck equation, we have from Ref.\ \cite{Lang69} that the
kinetic prefactor $\kappa$=$\beta\gamma|\omega_0|$, where $\gamma$ is
the fundamental rate of fluctuation.  With Eqs.\ (\ref{int-eq0}),
(\ref{acf-eq16}), and (\ref{acf-eq19}), this gives the nucleation rate
density
\begin{equation}
\frac\Gamma{NL} = 15\gamma(\sqrt{2J}N)^{-1/2}
\left(\case\beta\pi\right)^{3/2}
|\alpha|^{3/8}|\lambda|^{11/8}\left[1+O(\lambda^{1/2})\right]
e^{-\beta\Delta F}\;.\label{acf-eq20}
\end{equation}
As was pointed out earlier, for large $N$, unless $H$ is extremely
close to $H_{\rm s}$, the free-energy cost $\Delta F$ of surmounting
the nucleation barrier is large, so the exponential factor sets the
scale for the metastable lifetime.  However, for small $N$, or for
$H$$\approx$$H_{\rm s}$, the lifetime is more strongly dependent on
the dynamics and on the detailed structure of the saddle point.  In
Sec.\ \ref{sec-results} we will show how the crossover between these
two regimes depends on $N$ by deriving the finite-range-scaling
properties of ${\rm Im}\tilde f$.  In the same section we will also
directly compare numerical results from the transfer-matrix method
outlined in the next section with the exponential weight in Eq.\
(\ref{acf-eq19}), where the gap $\Delta F$ is determined numerically
from Eq.\ (\ref{mod-eq6}).

\section{The Constrained-Transfer-Matrix Method}
\typeout{The Constrained-Transfer-Matrix Method}
\label{sec-method}

Recently one of us \cite{Rikv89,Rikv91a} introduced a transfer-matrix
method, based on the concept of ``constrained'' joint probability
densities, to obtain analogues of the free-energy density for
constrained states.  Preliminary applications
\cite{Rikv89,Rikv91a,Rikv92a} to the Q1DI model have shown
qualitative agreement between the behavior of the free-energy-density
analogue associated with the metastable eigenvalue branch of the
transfer matrix and the analytically continued free-energy density.
In this study we use this constrained-transfer-matrix (CTM) method
with finite-range scaling \cite{Rikv93} to obtain more quantitative
results for the scaling of the imaginary part of the metastable
``constrained'' free-energy density.  In this section we outline the
method, and we describe the use of symmetry reductions in its
application to the Q1DI model.

The standard transfer-matrix method is usually applied as follows
\cite{Domb60}.  Starting with a Hamiltonian for an $N$$\times$$L$
$q$-state system that is invariant under translation in the
$L$-direction, such as Eq.\ (\ref{mod-eq0}), we write it as a sum of
layer Hamiltonians: ${\cal H}$=$\sum_{i=1}^L\bar{\cal
H}(x_i,x_{i+1})$.  The layer Hamiltonians $\bar{\cal H}$ are chosen to
depend only on the $q^{2N}$ configurations of a pair of adjacent
layers and are chosen so that the form of $\bar{\cal H}$ is
independent of the layer index.  We then define the $q^N$$\times$$q^N$
transfer matrix $\bf T$ as an operator in the dual space of
configurations
$|x_i\rangle$$\equiv$$|s_{i,1}\rangle\cdots|s_{i,N}\rangle$ of each of
the two adjacent layers:
\begin{equation}
\label{meth-eq0}
{\bf T} =
\sum_{x,x'}|x\rangle e^{-\beta\bar{\cal H}(x,x')}\langle x'|\;.
\end{equation}
The partition function for the $L$-layer system with periodic boundary
conditions is then $Z$=${\rm Tr}({\bf T}^L)$, which in the limit
$L$$\rightarrow$$\infty$ gives the free-energy density in terms of the
largest eigenvalue $\lambda_0$ of ${\bf T}$ as
$f$=$-(N\beta)^{-1}\ln\lambda_0$. (Here we index eigenvectors
$|\alpha\rangle$, $\alpha$=$0,1,2,\ldots$, in order of decreasing
magnitude of their respective eigenvalues:
$\lambda_0$$>$$|\lambda_1|$$\ge$$|\lambda_2|$$\ge$$\cdots$.)  Since
$\bf T$ is positive, $\lambda_0$ is positive and nondegenerate, and
the eigenvector $|0\rangle$ can be chosen to have all positive
elements by the Perron-Frobenius theorem \cite{Marc64}.  The
equilibrium state of the system is characterized by the joint and
marginal probability densities
\begin{eqnarray}
P_0(x_i,x_{i+k}) &=&
\langle0|x_i\rangle\langle x_i|(\lambda_0^{-1}
{\bf T})^{|k|}|x_{i+k}\rangle\langle x_{i+k}|0\rangle\nonumber\\
P_0(x_i) &=& \langle0|x_i\rangle\langle x_i|0\rangle\;.
\label{meth-eq1}
\end{eqnarray}

In the following discussion we restrict $\bf T$ to be symmetric.  (For
the Q1DI model $\bf T$ is made symmetric by symmetrizing the layer
Hamiltonian $\bar{\cal H}$.)  For each eigenvalue $\lambda_\alpha$ of
$\bf T$ we define a constrained transfer matrix (CTM) ${\bf T}_\alpha$
to commute with $\bf T$, so that it can be expanded in the
eigenvectors $|\alpha'\rangle$ of $\bf T$:
\begin{equation}
{\bf T}_\alpha =
\sum_{\alpha'}|\alpha'\rangle\mu_\alpha(\alpha')\langle\alpha'|\;.
\label{meth-eq2}
\end{equation}
The ``reweighted'' eigenvalues $\mu_\alpha$ are chosen to produce
``constrained'' probability densities,
\begin{eqnarray}
P_\alpha(x_i,x_{i+k}) &=&
\langle\alpha|x_i\rangle\langle x_i|(\lambda_\alpha^{-1}
{\bf T}_\alpha)^{|k|}|x_{i+k}\rangle\langle x_{i+k}|\alpha\rangle
\nonumber\\
P_\alpha(x_i) &=& \langle\alpha|x_i\rangle\langle x_i|\alpha\rangle\;,
\label{meth-eq3}
\end{eqnarray}
in analogy with the equilibrium ($\alpha$=0) case.  It was pointed out
in Refs.\ \cite{McCr78,Priv82a,Priv82b} that the constrained marginal
probability densities $P_\alpha(x)$, as defined above, can be
interpreted as actual probability densities of single-layer
configurations in a constrained state.  For example, the expectation
value $\langle M\rangle_\alpha$ of the layer magnetization density for
a constrained state $|\alpha\rangle$ is given by
\begin{equation}
\langle M\rangle_\alpha = \sum_xP_\alpha(x)M(x)\;.
\label{meth-eq4}
\end{equation}

In order to ensure that the entire system is characterized by
$P_\alpha(x)$, the matrix ${\bf T}_\alpha$ must be chosen so that the
constrained joint probability densities $P_\alpha(x,x')$ satisfy the
following regularity conditions: ({\em i\/}) that $P_\alpha(x)$ can be
obtained by summing over the configurations of one layer,
$P_\alpha(x)$=$\sum_{x'}P_\alpha(x,x')$; ({\em ii\/}) that
$P_\alpha(x_i,x_{i+k})$ is well-defined for $k$=0,
$P_\alpha(x_i,x'_i)$=$\delta_{x_i,x'_i}P_\alpha(x_i)$; and ({\em
iii\/}) that $P_\alpha(x_i,x_{i+k})$ reflects stochastic independence
in the limit $|k|$$\rightarrow$$\infty$, $\lim_{|k|\rightarrow\infty}
P_\alpha(x_i,x_{i+k})$=$P_\alpha(x_i)P_\alpha(x_{i+k})$.  For a matrix
${\bf T}_\alpha$ chosen to commute with $\bf T$, these requirements
are satisfied if ${\bf T}_\alpha$ has the same rank as $\bf T$, and if
its dominant eigenvalue is $\lambda_\alpha$.  The reweighting scheme
used in this work is
\begin{equation}
\mu_\alpha(\alpha')=\left\{\begin{array}{cl}
\lambda_\alpha^2/\lambda_{\alpha'}
& (\lambda_{\alpha'}>\lambda_\alpha)\\
\lambda_{\alpha'}
& (\lambda_{\alpha'}\leq\lambda_\alpha)\end{array}\right.
\label{meth-eq5}
\end{equation}
This choice ensures that any fluctuation described by
$P_{\alpha'}(x)$, with $\alpha'$$\ne$$\alpha$, decays away to the
$|\alpha\rangle$ eigenstate with a length scale
$|\ln(\lambda_\alpha/\lambda_{\alpha'})|^{-1}$.  Thus a fluctuation
with $\alpha'$$<$$\alpha$, which would otherwise grow, is effectively
suppressed.  We note, however, that the choice of ${\bf T}_\alpha$
given by Eqs.\ (\ref{meth-eq2}) and (\ref{meth-eq5}), which satisfies
the constraints, is not unique.  It is easy to see that we recover the
original transfer matrix $\bf T$ when we set $\alpha$=0.

The ``constrained'' free-energy density $f_\alpha$ is defined by
\begin{equation}
f_\alpha = \langle\bar{\cal H}\rangle_\alpha-\beta^{-1}S_\alpha\;,
\label{meth-eq6}
\end{equation}
where
\begin{equation}
\langle\bar{\cal H}\rangle_\alpha =
\frac1N\sum_{x_i,x_{i+1}}P_\alpha(x_i,x_{i+1})
\bar{\cal H}(x_i,x_{i+1})
\label{meth-eq7}
\end{equation}
is the expectation value of $\bar{\cal H}$ with respect to
$P_\alpha(x,x')$, and a generalized entropy $S_\alpha$ is defined
using $P_\alpha(x,x')$ in analogy with the source entropy of a
stationary, ergodic Markov chain (See {\it e.g.} Ref.\ \cite{Blah87}):
\begin{equation}
S_\alpha = -\frac1N\sum_{x_i,x_{i+1}}P_\alpha(x_i,x_{i+1}){\rm Ln}
\langle x_i|\lambda_\alpha^{-1}{\bf T}_\alpha|x_{i+1}\rangle\;.
\label{meth-eq8}
\end{equation}
The generalized free-energy density may be written in the form
\begin{equation}
f_\alpha = -\frac{\ln|\lambda_\alpha|}{\beta N}+\frac{1}{\beta N}
\sum_{x_i,x_{i+1}}P_\alpha(x_i,x_{i+1}){\rm Ln}
\frac{\langle x_i|{\bf T}_\alpha|x_{i+1}\rangle}
{\langle x_i|{\bf T}_0|x_{i+1}\rangle}\;.
\label{meth-eq9}
\end{equation}
The first term is analogous to the equilibrium case, and the second
term is complex-valued in general.  This is because ${\bf T}_\alpha$
is not a positive matrix in general.  To see this, note that for
$|\alpha\rangle$ ($\alpha$$\ne$0) to be orthogonal to $|0\rangle$, the
elements of $|\alpha\rangle$ must be of mixed sign.  Since
$\lambda_\alpha$ is the largest eigenvalue of ${\bf T}_\alpha$, the
largest contribution to ${\bf T}_\alpha$ is the projection
$|\alpha\rangle\lambda_\alpha\langle\alpha|$, which must contain
negative elements.  Therefore, the argument to the principal value of
the logarithm $({\rm Ln})$ may be negative.  We define the domain of
${\rm Ln}z$ to be $|z|$$>$0, $-\pi$$<$$\arg z$$\le$$\pi$, thus
choosing the branch cut along the negative real axis.  It is easy to
see that if $\alpha$=0, the second term of Eq.\ (\ref{meth-eq9})
vanishes, leaving the equilibrium free-energy density.

The above formalism is applicable to any system for which a symmetric
transfer matrix can be written.  A CTM study of the two-dimensional
Ising ferromagnet with nearest-neighbor interactions is reported in
Ref.\ \cite{Guen93}, and a similar study of a three-state model with
long-range interactions is reported in Ref.\ \cite{Fiig94}.  One
reason we have chosen to study the Q1DI model is that the Hamiltonian
is invariant under any permutation of spins in a given layer, and thus
the rank of the transfer matrix for this model is $N$+1 \cite{Novo86},
instead of the typical value of $2^N$ for short-range Ising systems.
We use the low rank of the transfer matrix to our advantage by
expressing the matrix in a reduced basis.  The computer time and
memory saved by this reduction allow us to study systems with very
large cross sections.

The reduction proceeds as follows.  Let $X$ be the $2^N$-dimensional
vector space in the basis $\{|x_j\rangle\}$ ($j$=$1,\ldots,2^N$) of
layer configurations, and let $G$ be the group of spin permutations of
the layer, which are represented by unitary operators ${\bf
U}_k$:$X$$\rightarrow$$X$ ($k$$\in$$G$).  This group partitions
$\{|x_j\rangle\}$ into $N$+1 equivalence classes ${\cal C}_n$
($n$=$0,\ldots,N$), each of which may naturally be associated with a
magnetization $m_n$.  This partition induces a decomposition of the
configuration space as $X$=$\bigoplus X_n$, where each $X_n$ is the
subspace spanned by the basis vectors $|x_j\rangle$$\in$${\cal C}_n$.
That $\bf T$ is $G$-invariant means that ($\forall k$$\in$$G$) ${\bf
TU}_k$=${\bf T}$, which implies the weaker property that $\bf T$
commutes with every ${\bf U}_k$, and thus may be simultaneously
diagonalized with any ${\bf U}_k$ by a unitary transformation ${\bf
S}_k\cdot{\bf S}_k^{-1}$. Thus if $\bf D_T$ and ${\bf D}_k$ are
diagonal matrices containing the eigenvalues of $\bf T$ and ${\bf
U}_k$ respectively, then ${\bf D_TD}_k$=${\bf S}_k{\bf TU}_k{\bf
S}_k^{-1}$=${\bf S}_k{\bf TS}_k^{-1}$=$\bf D_T$.  Since the
eigenvalues of ${\bf U}_k$ are roots of unity, it follows that for
each eigenvalue of ${\bf U}_k$ not equal to unity, the corresponding
transfer-matrix eigenvalue is zero.  Therefore, $\bf T$ has nonzero
eigenvalues only in the subspace of $X$ for which the eigenvalues of
${\bf U}_k$ are unity for every $k$$\in$$G$.  This is the
$N$+1-dimensional $G$-invariant subspace of $X$.

We construct a basis in this subspace as follows.  For each class
${\cal C}_n$, let $|m_n\rangle$ denote the normalized $G$-invariant
vector projecting only into $X_n$:
\begin{equation}
|m_n\rangle = g_n^{-1/2}\sum_{x_j\in{\cal C}_n}|x_j\rangle\;,
\label{meth-eq10}
\end{equation}
where $g_n$=${N\choose n}$ is the number of basis vectors in ${\cal
C}_n$.  Using Eqs.\ (\ref{meth-eq0}) and (\ref{meth-eq10}), we rewrite
the transfer matrix in this reduced basis as
\begin{eqnarray}
{\bf T} &=& \sum_{n,n'}\sum_{x\in{\cal C}_n}\sum_{x'\in{\cal C}_{n'}}
|x\rangle e^{-\beta\bar{\cal H}(x,x')}\langle x'|\nonumber\\
&=& \sum_{n,n'}|m_n\rangle (g_ng_{n'})^{1/2}
e^{-\beta\bar{\cal H}(m_n,m_{n'})}\langle m_{n'}|\;.
\label{meth-eq11}
\end{eqnarray}
We can similarly decompose the sums over configurations present in
Eqs.\ (\ref{meth-eq4}) and (\ref{meth-eq7}--\ref{meth-eq9}) into sums
over the reduced basis vectors \cite{Rikv89}.  The transfer-matrix
results presented in the next section were obtained using this reduced
representation.

\section{Numerical Results}
\typeout{Numerical Results}
\label{sec-results}

In this section we present the results of two numerical methods used
to study the properties of the metastable phase in the Q1DI model.  We
numerically integrate the Euler-Lagrange equation, in the discrete
form of Eq.\ (\ref{mod-eq6}), to estimate the free-energy cost of
nucleation, which we use to verify the field-theoretical results of
Sec.\ \ref{sec-analytic}.  We also apply the CTM formalism outlined in
Sec.\ \ref{sec-method} to the model.  We perform finite-range scaling
on the constrained free-energy density $f_\alpha$ of Eq.\
(\ref{meth-eq9}), and we compare the results of the two numerical
methods and the theoretical predictions of Sec.\ \ref{sec-analytic}.

\subsection{Numerical Integration Results}
\label{sec-resni}

To solve Eq.\ (\ref{mod-eq6}) numerically, we represent the field
$\phi$ as an $L$-vector in which each element is allowed to vary
continuously.  For the numerical solutions presented in this work we
have chosen $L$=100.  (Doubling $L$ in this case produced no
discernible change in the results for the fields and temperatures we
studied.)  We set the interaction constants $J$=1/2 and $J'$=0, so
that $\beta_c$=1, and we obtained numerical results for temperatures
in the range $0.1T_{\rm c}$$\le$$T$$\le$$0.8T_{\rm c}$.  Setting
$H$$>$0, we determine the magnetization density $\phi_{\rm core}$ at
the core of the droplet by solving $f_{\rm CW}(\phi_{\rm
core})$=$f_{\rm CW}(\phi_{\rm ms})$ under the constraint $\phi_{\rm
core}$$>$$\phi_{\rm ms}$.  We then force the type-II solution by
setting $\phi_1$=$\phi_{\rm core}$, and $\phi_{L/2}$=$\phi_{\rm ms}$
as boundary conditions.  The layers between $L/2$ and $L$ are given by
the symmetry imposed by periodic boundary conditions.

The methods chosen to solve Eq.\ (\ref{mod-eq6}) are the well-known
shooting and relaxation methods. (See, {\it e.g.}\ Ref.\
\cite{Pres92}.)  Taking the explicit form of Eq.\ (\ref{mod-eq5a}) for
$f_{\rm CW}$, we use Eq.\ (\ref{mod-eq6}) directly to shoot from
$i$=1,
\begin{equation}
\phi_{i+1}=-\phi_{i-1}-H/J+(\beta J)^{-1}\tanh^{-1}\phi_i\;,
\label{resa-eq0}
\end{equation}
stopping before the solution becomes catastrophic.  The remainder of
the system is given the value $\phi_{\rm ms}$, and a relaxation method
is employed, using a variant of Eq.\ (\ref{resa-eq0}),
\begin{equation}
\phi_i = \tanh\left(
\beta\left[H+\case12J(\phi_{i-1}+\phi_{i+1})\right]\right)\;.
\label{resa-eq1}
\end{equation}
The values are then relaxed iteratively in a checkerboard fashion,
each layer being adjusted by an increment $\Delta\phi_i$, until
$\sum_i|\Delta\phi_i|$$<$$10^{-6}$ for one iteration.  The free-energy
cost $\Delta F$ of forming the droplet is then obtained using Eq.\
(\ref{mod-eq5}):
\begin{equation}
\frac{\Delta F}N = \sum_{i=1}^L\left[
\case14(\phi_{i+1}-\phi_i)^2
+f_{\rm CW}(\phi_i)-f_{\rm CW}(\phi_{\rm ms})\right]\;.
\label{resa-eq2}
\end{equation}

In Fig.\ \ref{fig2} the numerical solutions for various temperatures
are shown as functions of $|\lambda|/H_{\rm s}=(H_{\rm s}-H)/H_{\rm
s}$ and are compared with the theoretical result for $\Delta F$ from
Eq.\ (\ref{acf-eq8}). Corrections to the $\phi^3$ field theory are
seen to be substantial, especially for low temperatures, except for a
region very close to the spinodal.  We can account for this rapid
departure of the numerical solutions from the $\phi^3$ field theory by
considering the leading field-dependent corrections to the cubic
potential used in Eq.\ (\ref{acf-eq0}).  We can write the first
correction to the integrand of Eq.\ (\ref{acf-eq0}) as
$\frac14\epsilon v^4$, where
\begin{equation}
\epsilon = \frac16\left.
\frac{\partial^4f_{\rm CW}}{\partial\phi^4}\right|_{\phi_s}
= \frac{\beta^2}{\beta_{\rm c}^3}
\left(\frac43-\frac{\beta_{\rm c}}\beta\right)\;.
\end{equation}
To leading order, the correction to the LHS of Eq.\ (\ref{acf-eq1}) is
then $\epsilon v^3$, and Eq.\ (\ref{acf-eq5}) may be rewritten
\begin{equation}
\Delta{\cal F} =
-\lambda\phi_{\rm s}-\case23\alpha v_0^3-\case34\epsilon v_0^4
-\frac1L\int_0^L\left[
\case16\alpha u^3+\case14\epsilon(2v_0u^3+u^4)\right]dr\;.
\label{resa-eq3}
\end{equation}
The length scale of the droplet is still given approximately by
$\xi_{\rm r}$, so performing the integration in the last term
introduces a factor of $\xi_{\rm r}$.  The first correction to $\Delta
F$ thus has the temperature and field dependence of $\epsilon
v_0^4\xi_{\rm r}$$\sim$$\epsilon|\alpha|^{-9/4}|\lambda|^{7/4}$.

We can verify the presence and form of this correction by studying the
behavior of the numerically obtained $\Delta F$ as
$|\lambda|$$\rightarrow$0.  For a given $\beta$, we generate a
sequence of field deviations $|\lambda_n|$=$\lambda_0/n^2$ (not to be
confused with the transfer-matrix eigenvalues), where $\lambda_0$ is a
constant, for an increasing sequence of integers $n$.  We then
calculate the free-energy cost of the critical droplet $\Delta F_n$,
using the numerical integration procedure outlined above, for a field
$H = H_{\rm s}-\lambda_n$.  If we define two sequences $P_n$=$\Delta
F_n/\Delta F$, where $\Delta F$ is the analytic result of Eq.\
(\ref{acf-eq8}) at $\lambda$=$\lambda_n$ with no corrections assumed,
and $Q_n = n(P_n-1)$, then the dominant contributions to $\Delta F_n$
should give the following expansions:
\begin{eqnarray}
P_n &=& 1+A_1(\beta)n^{-1}+A_2(\beta)n^{-2}+\cdots\nonumber\\
Q_n &=& A_1(\beta)+A_2(\beta)n^{-1}+\cdots\;,
\end{eqnarray}
where the amplitudes $A_i(\beta)$ are functions only of $\beta$, and
in particular, $A_1(\beta)$$\propto$$\epsilon|\alpha|^{-3/2}$.  For
even $n$, we transform the numerical sequences $P_n$ and $Q_n$ in
order to remove the first corrections.  The method we use, a variant
of Neville-Aitken extrapolation \cite{Gutt89}, is described in detail
in Ref.\ \cite{Rikv93}, where it was used to extrapolate with much
success finite-size data carrying very substantial corrections.  We
apply it here to remove $O(n^{-1})$ corrections to $P_n$ and $Q_n$:
\begin{eqnarray}
P'_n &=& 2P_n-P_{n/2}\nonumber\\
Q'_n &=& 2Q_n-Q_{n/2}\;,
\end{eqnarray}
which give $P'_n=1+O(n^{-2})$ and $Q'_n=A_1(\beta)+O(n^{-2})$.  The
constancy of the leading terms in $P_n$ and $Q_n$ was verified by
calculating estimates $W^P_n$ and $W^Q_n$ to the exponents of $n$ in
these terms, which we expect to be zero:
\begin{equation}
W^P_n=(\ln2)^{-1}\ln(P'_n/P'_{n/2})\;,
\end{equation}
and similarly for $Q_n$, which gives the leading exponent to
$O(n^{-2})$.  For each temperature, we computed the sequences $\Delta
F_n$ using $\lambda_0$=$0.5H_{\rm s}$ and $n$=$1,2,3,\ldots$.  The
sequences were terminated when $P_n$ came within 0.5\% of unity.  The
resulting estimates of the $|\lambda|$-exponent ($W^P_n$+$\case54$) in
the leading term lay between 1.249 and 1.251, and the estimates for
the $|\lambda|$-exponent in the first correction term
($\case12W^Q_n$+$\case74$) lay between 1.74 and 1.75, both confirming
the field-theoretical predictions of 5/4 and 7/4, respectively
\cite{GormXX}.  We used the value of $P'_n$ as an estimate for the
leading coefficient to $P_n$.  These estimates lay between 0.9998 and
1.0001, confirming Eq.\ (\ref{acf-eq8}) to high accuracy.  Dividing
the predicted temperature dependence of $A_1$ from $Q'_n$, we found
the resulting coefficient to vary only about 15\% over the entire
range of temperatures, giving evidence that the predicted temperature
dependence of the first correction to $\Delta F$ is also as predicted.
For $H/H_{\rm s}$$\ll$1, the logarithmic cusp singularity in $\Delta
F$ predicted in Refs.\ \cite{Lang67,Buet81} was also observed in our
numerical integration results \cite{GormXX}.  Since transfer-matrix
results were not obtainable at sufficiently low fields, for this work
we will not discuss this field region further.

\subsection{Constrained-Transfer-Matrix Results}
\label{sec-resctm}

The CTM method was applied to Q1DI systems of finite cross section $N$
and infinite length, with interaction constants $J$=1/2 and $J'$=0, so
that $\beta_{\rm c}$=1.  The temperatures studied using the CTM method
were in the range $0.2T_{\rm c}$$\le$$T$$\le$$0.8T_{\rm c}$, and data
at each temperature was collected for fields ranging from $H$=0 to
$H$$\approx$$1.2H_{\rm s}$.  The matrix was tridiagonalized by a
Householder reduction, and then diagonalized by a QL algorithm with
implicit shifts \cite{Pres92}.  The computations were performed in
128-bit precision on a Cray Y-MP 4/32 using about 200 CPU-hours.
Computations were limited mainly by machine precision, since the
process described in Sec.\ \ref{sec-method} involves the reweighting
and resumming of eigenvalues whose magnitudes range over thirty or
more decades.  The lower limit on the temperatures studied was due to
limited machine precision, whereas the upper limit was due to the
extremely large cross sections required to minimize finite-size
effects.  For most temperatures, systems were studied for $N$ up to
the range 100$<$$N$$<$200, although for the highest temperatures we
were able to obtain results for $N$$\approx$500.

In Fig.\ \ref{fig3} typical eigenvalue and layer magnetization spectra
of the transfer matrix for the Q1DI model are shown as functions of
$H$.  The eigenvalue spectrum is shown on a scale so that its lowest
branch is the equilibrium free-energy density of the system.  This
spectrum can be divided into three regions \cite{Novo86}.  The first
is a region of low-lying states with positive magnetization.  Above
this region is a region of alternately polarized states that become
nearly degenerate at regular intervals in $H$.  Since all the
eigenvectors of the reduced transfer matrix belong to the same
symmetry class, the eigenvalue branches are never actually degenerate
and therefore do not cross, but rather quickly reverse their
direction.  In Ref.\ \cite{Novo86} the gaps between these nearly
degenerate eigenvalues were found to approach zero exponentially with
$N$.  The quick reversal of the eigenvalue branches can be seen in the
near discontinuities in the magnetization branches, in which the
values of the magnetizations associated with the two nearly degenerate
eigenvalues are effectively traded.  This region terminates
approximately at the Curie-Weiss spinodal field $H_{\rm s}$ of Eq.\
(\ref{mod-eq9a}).  The third region, above the second, consists of
eigenvalues that have little polarization.  For the results presented
here we are concerned with the lowest-lying states of the second
region, that is, the lowest-lying states with magnetization opposite
the applied magnetic field.  These states form a composite branch that
agrees well with the values shown for the free-energy density and
magnetization of the metastable state for the Curie-Weiss ferromagnet
in the thermodynamic limit.  This agreement improves as $N$ is
increased.

The real and imaginary parts of $f_\alpha$ from the CTM, as computed
by Eq.\ (\ref{meth-eq9}), are shown in Fig.\ \ref{fig4} for a single
system size at fixed temperature.  The real part of $f_\alpha$
exhibits a composite branch that is nearly identical to the composite
``metastable'' branch shown in Fig.\ \ref{fig3}, and thus is also in
good agreement with the known free-energy density of the metastable
state.  In addition, the corresponding values of $|{\rm Im}f_\alpha|$
for this branch are nearly zero except in the region of the spinodal.
Figure \ref{fig5} shows the same branches of $|{\rm Im}f_\alpha|$ as
Fig.\ \ref{fig4} on a logarithmic scale. Note that the values of
$|{\rm Im}f_\alpha|$ range over more than 30 decades.  The structure
of the ``metastable'' branch as a succession of lobes rather than as a
smooth function is due to the mixing of nearly-degenerate eigenvectors
(see Fig.\ \ref{fig3}).  Since we were interested in obtaining values
from branches representing the single ``metastable'' phase
$|\alpha\rangle$, we selected values of $H$ for which
$\lambda_\alpha$=$(\lambda_{\alpha+1}\lambda_{\alpha-1})^{1/2}$, thus
ensuring a ``safe'' distance from the near-degeneracies.  The
resulting points are shown on the figure and together roughly give
tangent points for a reasonable envelope for the lobes.  We identify
this envelope with ${\rm Im}f_{\rm ms}$, the imaginary part of the
constrained free-energy density for the metastable state.  The
envelope for a larger system at the same temperature is also shown.
By inspecting the envelopes for many system sizes, we observed roughly
an exponential dependence on $N$ well inside the spinodal and a
crossover to a slower scaling near the spinodal.  The evidence for
this scaling behavior will be made clear and quantitative in the next
subsection.

The presence of the crossover can be seen clearly in Fig.\ \ref{fig6},
where the second derivative of $|{\rm Im}f_{\rm ms}|$ with respect to
$H$ is plotted versus reduced field $H/H_{\rm s}$ for a fixed
temperature and various values of $N$.  Each derivative was calculated
by the midpoint method, using only the ``tangent'' points chosen as
described above.  As $N$ increases, a singularity in this derivative
develops and pushes closer to the spinodal.  Although the value of
$H_{\rm s}$ is temperature dependent, we observed the same behavior in
$|{\rm Im}f_{\rm ms}|$ for the entire range of temperatures studied.

\subsection{Finite-Range-Scaling Results}
\label{sec-resfrs}

First we turn our attention to the scaling behavior of $|{\rm
Im}f_{\rm ms}|$ for fields in the range 0$<$$H$$<$$H_{\rm s}$, and we
consider fields far enough from the spinodal that the Boltzmann weight
$e^{-\beta\Delta F}$ sets the scale for the metastable lifetime.
Since the transfer-matrix data lack the extremely high resolution
required to determine the prefactor to this weight, we concentrate on
the quantity $\ln|{\rm Im}f_{\rm ms}|$.  Assuming a form for $|{\rm
Im}f_{\rm ms}|$ as given by Eqs.\ (\ref{acf-eq8}) and
(\ref{acf-eq19}), we write
\begin{equation}
\ln|{\rm Im}f_{{\rm ms},N}(T,H)| \sim -\beta N^\sigma\Delta(T,H)\;,
\label{resc-eq0}
\end{equation}
where the exponent $\sigma$ and the function $\Delta(T,H)$ are
undetermined.

For each system size $N$ with $N$$\equiv$0(mod 4), we calculate
finite-range estimates for $\sigma$ at fixed $T$ and $H$ assuming that
the dominant correction is from the prefactor and is therefore $O(\ln
N)$:
\begin{equation}
\sigma_N = (\ln2)^{-1}\ln\left(
\frac{\ln|{\rm Im}f_{{\rm ms},N}|-\ln|{\rm Im}f_{{\rm ms},N/2}|}
{\ln|{\rm Im}f_{{\rm ms},N/2}|-\ln|{\rm Im}f_{{\rm ms},N/4}|}
\right)\;,\label{resc-eq1}
\end{equation}
which gives $\sigma_N=\sigma+O(N^{-1})$ \cite{Rikv93}.  Whereas the
values for $\ln|{\rm Im}f_{{\rm ms},N}|$ in Eq.\ (\ref{resc-eq1}) were
taken directly from the tangent points, the values of $\ln|{\rm
Im}f_{\rm ms}|$ for smaller systems had to be interpolated to the same
field value since both the number and positions of the lobes are
$N$-dependent.  We performed the interpolation by fitting a quadratic
form exactly to the three nearest points.  Figure \ref{fig7} shows
$\sigma_N$ for the largest values of $N$ numerically attainable,
plotted versus reduced field for various temperatures.  Clearly these
estimates are quite consistent with the $N$-dependence ($\sigma$=1) of
$\Delta F$ given by Eq.\ (\ref{acf-eq8}).  Note that as the estimates
approach $H_{\rm s}$, they drop significantly, indicating the rapidly
decreasing importance of the $N$-dependent exponential factor of
$|{\rm Im}f_{{\rm ms},N}|$ compared to the algebraic prefactor, as was
pointed out in Sec.\ \ref{sec-analytic}.

For even $N$, finite-$N$ estimates for $\Delta(T,H)$ were calculated
assuming that $\sigma$=1, and again that the dominant correction is
$O(\ln N)$:
\begin{equation}
\Delta_N = -\case2{\beta N}
(\ln|{\rm Im}f_{{\rm ms},N}|-\ln|{\rm Im}f_{{\rm ms},N/2}|)\;.
\label{resc-eq2}
\end{equation}
Values for $\ln|{\rm Im}f_{{\rm ms},N}|$ were taken as described in
the previous paragraph.  Figure \ref{fig8} shows a typical set of
estimates for $\Delta F$.  Due to logarithmic corrections, the
convergence of naive estimates using Eq.\ (\ref{resc-eq0}) directly,
$\Delta(T,H)$$\approx$$(\beta N)^{-1}\ln|{\rm Im}f_{\rm ms}|$, is
quite slow.  Using Eq.\ (\ref{resc-eq2}) gives greatly improved
estimates, which agree much better with the free-energy cost of
nucleation obtained by numerical integration, and the range of fields
over which there is good agreement extends much closer to the
spinodal.  In addition, these extrapolated estimates show a trend near
the spinodal toward the numerical interation result, although the
limited accuracy in the interpolations used precluded any further
extrapolation.

Figure \ref{fig9} shows $\Delta_N$ for the largest values of $N$
numerically attainable, plotted versus reduced field for various
temperatures.  These are compared with the free-energy cost of
nucleation obtained though Eq.\ (\ref{resa-eq2}) by numerically
integrating the Euler-Lagrange equation as described in Sec.\
\ref{sec-resni}.  As can be seen from Figs.\ \ref{fig8} and
\ref{fig9}, the agreement between the extrapolated CTM estimates and
the exact results is impressive.  The relative error for these
estimates were consistently below 2\% over most of the field and
temperature range and remained below 5\% until the crossover region
was entered.  By comparing Fig.\ \ref{fig9} with Fig.\ \ref{fig2},
where the free-energy cost of nucleation obtained by numerical
integration is shown together with the $\phi^3$ field-theoretical
results, one sees that the CTM estimates for $\Delta F$ faithfully
reproduce the higher-order corrections to the $\phi^3$ field theory
over a wide range of fields and temperatures.  This strongly indicates
the consistency between the CTM method and the droplet theory of
nucleation, at least whenever a substantial free-energy barrier
against nucleation exists.

Next we consider the behavior of $|{\rm Im}f_{\rm ms}|$ near the
spinodal, where the extrapolated CTM estimates $\Delta_N$ lie
significantly above the infinite-$N$ free-energy cost $\Delta F$
obtained by numerical integration.  As discussed in Sec.\
\ref{sec-analytic}, here we must carefully consider the behavior of
the prefactor to the Boltzmann weight.  Since the mean-field spinodal
is a line of critical points in a $\phi^3$ field theory, we can apply
critical finite-range scaling \cite{Rikv93} in this region.  By
recasting Eqs.\ (\ref{acf-eq6}), (\ref{acf-eq8}), and (\ref{acf-eq19})
in terms of a scaling variable $\zeta$=$R_N^{4/5}|\lambda|$, we
determine the finite-range scaling relation for $|{\rm Im}\tilde f|$
to be
\begin{equation}
|{\rm Im}\tilde f| = R_N^{-6/5}\Phi(\beta,\zeta)\;,\label{resc-eq3}
\end{equation}
where the scaling function $\Phi$ depends on the sign of
$\lambda=H-H_{\rm s}$:
\begin{equation}
\Phi(\beta,\zeta) = \left\{\begin{array}{ll}
12(\sqrt2\pi\beta)^{-1/2}|\alpha|^{-1/8}\zeta^{7/8}
\exp\left(-\case{24}5\sqrt2\beta|\alpha|^{-3/4}\zeta^{5/4}\right)
& (|H|<|H_{\rm s}|)\\
\case23|\alpha|^{-1/2}\zeta^{3/2}
& (|H|>|H_{\rm s}|)\;.\end{array}\right.\label{resc-eq4}
\end{equation}
Note that as $\zeta$$\rightarrow$0 from either direction, $\Phi$ goes
to zero, with a cusp singularity at $|H|$=$|H_{\rm s}|^-$.  This
singular scaling function cannot be reproduced for finite $N$.  We
therefore expect that the finite-$N$ CTM estimates for $|{\rm
Im}f_{\rm ms}|$ near $H_{\rm s}$ should be dominated by corrections to
scaling.  Indeed, the transfer-matrix results for $|{\rm Im}f_{\rm
ms}|$ only showed qualitative agreement with the field theory for
$H$$\approx$$H_{\rm s}$.  Power-law scaling was found at $H_{\rm s}$,
but the exponent of $R_N$ varied between roughly $-0.8$ at low
temperatures and $-1$ at high temperatures, in significant
disagreement with Eq.\ (\ref{resc-eq3}), and large fluctuations in the
data precluded the use of extrapolation techniques.  In addition, the
values of the scaling functions calculated from $|{\rm Im}f_{\rm ms}|$
were consistently greater than those of the theoretical scaling
functions, varying from a factor of roughly 2 at low temperatures to 5
at high temperatures.  Several possible explanations for these results
will be considered in the next section.

\section{Discussion}
\typeout{Discussion}
\label{sec-disc}

In this work we have applied a scalar field theory and the recently
developed constrained-transfer-matrix (CTM) method to the study of
metastability in the ferromagnetic quasi-one-dimensional Ising model,
a two-state model with weak, long-range interactions.  By solving the
Euler-Lagrange equation associated with the Ginzburg-Landau-Wilson
Hamiltonian for the model, we have identified the equilibrium and
metastable states as spatially uniform configurations with the order
parameter taking the Curie-Weiss mean-field values, and we have
identified the critical fluctuation through which the metastable state
decays.  By computing the free-energy cost of the critical fluctuation
numerically, we were able to calculate the ``Boltzmann weight'' that
appears in the analytic continuation of the free energy across the
first-order phase transition for the entire region of metastability.
Except in the region of the classical spinodal, this weight gives the
dominant time scale for decay of the metastable state.  In the region
of the spinodal we have mapped the Hamiltonian to a $\phi^3$ field
theory to obtain an expression for the analytically continued free
energy with no undetermined parameters.  Assuming the applicability of
the Fokker-Planck equation to the dynamics, we have thus also obtained
an exact expression for the decay rate of the metastable state for
this model near the classical spinodal.  In performing the analytic
continuation of the free energy, we have found that its corrections
are quite substantial, leading to large differences in, for example,
the free-energy cost of nucleation as one moves away from the spinodal
toward the first-order transition.  Numerical solution of the
Euler-Lagrange equation strongly supports the field-theoretical result
for the free energy-cost of nucleation, including the first correction
term.

We have outlined a method by which a complex analogue of the free
energy for a constrained system is obtained directly from the transfer
matrix for the unconstrained system.  Extensive symmetry reductions
have enabled us to collect transfer-matrix data for systems with very
large cross sections, and hence very large interaction ranges.  We
found that the real part of the constrained free-energy density
associated with the metastable phase, ${\rm Re}f_{\rm ms}$, rapidly
approaches the free-energy density of the metastable phase as
$N$$\rightarrow$$\infty$.  The associated imaginary part $|{\rm
Im}f_{\rm ms}|$ is extremely small, showing exponential dependence on
the interaction range over most of the region of metastability and a
crossover to power-law scaling near the classical spinodal.  We found
strong evidence for this crossover in the behavior of the second field
derivative of $|{\rm Im}f_{\rm ms}|$.  Using numerical extrapolation
techniques, we have demonstrated that over a wide range of fields and
temperatures within the region of metastability, the complex
``constrained'' free-energy density obtained by this method agrees
very well with the behavior of the analytically continued free-energy
density.  The estimated free-energy cost of the critical droplet was
found to lie within 2\% of the value obtained by solving the
Euler-Lagrange equation numerically.  In fact, for low temperatures
and for fields sufficiently far from the spinodal, the transfer-matrix
data were found to agree with the numerical solution much more closely
than the analytic field-theoretical results.

However, we found that the constrained free-energy density, as defined
by Eq.\ (\ref{meth-eq9}), using the reweighting scheme defined in Eq.\
(\ref{meth-eq5}), does not show consistent finite-range scaling at the
spinodal.  There are several possible explanations for this result.
If we consider the possibility that finite-size corrections to the
free-energy cost $\Delta F$ of nucleation are affecting the exponents
in the prefactor of $|{\rm Im}\tilde f|$, then those corrections must
be $O(\ln N)$ \cite{GormXX}.  (This rules out, for example, the effect
of relaxing Stirling's approximation, which was used to derive Eq.\
(\ref{mod-eq5}), since it gives only a correction of order unity.)
Another possibility is that the effects of eigenvector mixing are too
strong in this region to obtain reasonable estimates for $|{\rm
Im}f_{\rm ms}|$.  In addition, for finite cross sections $N$, the
value of the spinodal may not be entirely real, so that a dominant
contribution to $|{\rm Im}f_{\rm ms}|$ emerges from the CTM data due
to finite-size rounding.  If $f_{\rm ms}$ does represent the analytic
continuation of the free-energy density, then it is possible that the
eigenvalues are being reweighted incorrectly in this region, allowing
unwanted or unphysical fluctuations.  These and other possibilities
will be explored in further work.

The CTM method clearly shows promise in the characterization of
metastable phases.  As a nonperturbative method, it treats all
possible fluctuations in a single calculation, and hence does not
require any particular assumptions about the detailed structure of the
partition function.  Since it automatically identifies the
fluctuations that are important to nucleation, the CTM method has
potential applications to the study of metastability in, for example,
disordered systems, in which critical fluctuations are difficult to
characterize.  The usefulness of this method has been demonstrated in
a study of the 2D Ising ferromagnet \cite{Guen93}, in which excellent
quantitative agreement was found between field-theoretical and CTM
estimates of the surface free energy and the shape of the critical
droplets, as well as consistency with Monte-Carlo estimates of the
metastable lifetime \cite{Rikv94,Guen94}.  Furthermore, in a
three-state model with weak, long-range interactions \cite{Fiig94},
$|{\rm Im}f_{\rm ms}|$ appears to remain consistent with the
metastable lifetime, even under conditions of competing metastable
states, where it has been argued \cite{Gave89} that the analytic
continuation of the free energy may no longer be a valid measure of
the lifetime.  A Monte-Carlo study of metastability in the Q1DI model,
using recently developed techniques, is in progress, and a further
investigation of the connection between the dynamics of metastable
states and the transfer matrix is planned.

\acknowledgements
We would like to thank T.~Fiig and C.~C.~A. G\"unther for useful
discussions.  This work was supported in part by Florida State
University through the Supercomputer Computations Research Institute
(Department of Energy Contract No.\ DE-FC05-85ER25000), through the
Center for Materials Research and Technology, and through Cray Y-MP
supercomputer time, and by National Science Foundation Grant No.\
DMR-9013107.

\begin{figure}
\caption{Schematic diagrams of the solutions to Eq.\
(\protect\ref{mod-eq8a}) for the cases of $H$=0 and $H$$>$0.  The
solutions are shown together with sketches of the corresponding
Curie-Weiss free-energy density $f_{\rm CW}$.  The diamonds mark the
uniform (type-I) solutions.  The dashed horizontal line in each case
represents the interval over which the order parameter ranges in the
type-II solution.  The hashed regions represent the bands of type-III
solutions, which oscillate in space between magnetization densities
with the same value of $f_{\rm CW}$.  Below these sketches of the
intervals are sketches in real space of the type-II solutions.}
\label{fig1}
\end{figure}

\begin{figure}
\caption{Estimates for the free-energy cost $\Delta F$ of forming the
critical droplet in the Q1DI model, shown as a cross-section ($N$-)
density and as a function of the $H$-field deviation from the
spinodal, $|\lambda|/H_{\rm s} = (H_{\rm s}-H)/H_{\rm s}$.  The solid
curves are the results of numerically solving Eq.\
(\protect\ref{mod-eq6}) at temperatures $T/T_{\rm c}$=0.2, 0.4, 0.6,
and 0.8.  The curves corresponding to this order run from top to
bottom.  The dashed straight lines show the estimates from the
$\phi^3$ field-theoretical result of Eq.\ (\protect\ref{acf-eq8}) at
the same temperatures.  These lines appear in the same order, with the
two top lines nearly coincident.}
\label{fig2}
\end{figure}

\begin{figure}
\caption{Eigenvalue (top) and magnetization (bottom) spectra computed
from the transfer matrix for a 35$\times$$\infty$ Q1DI system at
$T$=$0.5T_{\rm c}$ and plotted versus $H$.  The value of the
Curie-Weiss spinodal field $H_{\rm s}$ is marked by the dotted
vertical line, and the values of the free-energy density and
magnetization for the equilibrium, metastable, and unstable stationary
states of the Curie-Weiss ferromagnet are marked by the thick gray
lines.  Fifteen eigenvalue branches lie above the chosen plotting
range, so they are not visible.  See Sec.\ \protect\ref{sec-resctm}
for a detailed description.}
\label{fig3}
\end{figure}

\begin{figure}
\caption{All branches of ${\rm Re}f_\alpha$ (top) and $|{\rm
Im}f_\alpha|$ (bottom) computed from constrained transfer matrices for
a 35$\times$$\infty$ Q1DI system at $T$=$0.5T_{\rm c}$ and plotted
versus $H$.  The dotted vertical line indicates the spinodal field.
The branches of ${\rm Re}f_\alpha$ are plotted on the same scale as
the eigenvalue branches of Fig.\ \protect\ref{fig3}, so that the two
spectra may be compared directly.}
\label{fig4}
\end{figure}

\begin{figure}
\caption{Branches of $\ln|{\rm Im}f_\alpha|$ computed from constrained
transfer matrices for a 35$\times$$\infty$ Q1DI system at
$T$=$0.5T_{\rm c}$ and plotted versus $H$.  The (+) symbols touching
the lobes are calculated at fields chosen by the criterion
$\lambda_\alpha$=$(\lambda_{\alpha+1}\lambda_{\alpha-1})^{1/2}$ for
the ``metastable'' eigenvector $|\alpha\rangle$.  These points are
joined by straight lines.  A second set of points, for $N$=70, is
shown to illustrate the $N$-scaling behavior of $\ln|{\rm
Im}f_\alpha|$.  This second set is expected to continue downward at
lower fields roughly along the dashed line, but the numerical
precision was not sufficient to resolve points in the low-field
region.  The spinodal field is marked by the dotted vertical line.}
\label{fig5}
\end{figure}

\begin{figure}
\caption{The second derivative of $|{\rm Im}f_\alpha|$ with respect to
$H$ for Q1DI systems with various cross sections $N$ at $T$=$0.2T_{\rm
c}$.  The symbols are the results of the midpoint method applied to
the points used to define the envelope over the lobes of $|{\rm
Im}f_\alpha|$. (See Fig.\ \protect\ref{fig5}.)  The lines joining
these points are splines serving only as guides to the eye.  The
dotted vertical line indicates the spinodal field.}
\label{fig6}
\end{figure}

\begin{figure}
\caption{Finite-$N$ estimates for the exponent $\sigma$ in Eq.\
(\protect\ref{resc-eq0}), for the largest $N$ attainable, plotted
against reduced field $H/H_{\rm s}$ for various temperatures $T$.  The
dotted horizontal line indicates the field-theoretical value
($\sigma$=1) over this range.}
\label{fig7}
\end{figure}

\begin{figure}
\caption{Finite-$N$ estimates for $\Delta(T,H)$ in Eq.\
(\protect\ref{resc-eq0}) with $\sigma$=1 fixed, plotted against
$|\lambda|/H_{\rm s}=(H_{\rm s}-H)/H_{\rm s}$ for $T$=$0.5T_{\rm c}$.
Each thin line connects the naive estimates one would obtain for a
particular $N$ directly from Eq.\ (\protect\ref{resc-eq0}).  These
estimates decrease in value as $N$ increases.  The data points are
estimates using Eq.\ (\protect\ref{resc-eq2}).  The diamonds are
results using the largest $N$ attainable.  The circles are results for
half of the largest $N$ and are plotted only in the region where
finite-size effects are large.  The thick line indicates the
free-energy cost of forming the numerical ``critical droplet''
solution of Eq.\ (\protect\ref{mod-eq6}), obtained by numerical
integration as described in Sec.\ \protect\ref{sec-resni}.}
\label{fig8}
\end{figure}

\begin{figure}
\caption{Finite-$N$ estimates for $\Delta(T,H)$ in Eq.\
(\protect\ref{resc-eq0}) with $\sigma$=1 fixed, extrapolated by Eq.\
(\protect\ref{resc-eq2}) for the largest $N$ attainable, and plotted
against $|\lambda|/H_{\rm s} = (H_{\rm s}-H)/H_{\rm s}$ for
various temperatures $T$.  The lines indicate the free-energy cost of
forming the numerical ``critical droplet'' solution of Eq.\
(\protect\ref{mod-eq6}) at the same temperatures, obtained by
numerical integration as described in Sec.\ \protect\ref{sec-resni}.
Compare with Fig.\ \protect\ref{fig2}.}
\label{fig9}
\end{figure}


\begin{thebibliography}{10}
\bibitem{frsam-note1}
The expression for $\nu_1$ was misprinted in Eq.\ A2 of Ref.\ \cite{Buet81}.
\bibitem{frsam-note2}
The results of Ref.\ \cite{Unge84} should be corrected as follows: (1) the
  exponent 1/4 in Eq.\ 4.8 should be $-1/4$; (2) the exponent 3$d$/8 in Eq.\
  4.11 should be $d$(6$-d$)/8; and consequently (3) the exponent (1/2+5$d$/8)
  in Eq.\ 4.16 should be $d$(1$-d$/8).
\end{thebibliography}
\end{document}